Regulation by Progestins, Corticosteroids and RU486 of Activation of Elephant Shark and Human Progesterone Receptors: An Evolutionary Perspective

Xiaozhi Lin[1], Wataru Takagi[2], Susumu Hyodo[2],
Shigeho Ijiri[3], Yoshinao Katsu[1, 4, *], Michael E. Baker[5, 6, *]

[1]Graduate School of Life Science, Hokkaido University, Sapporo, Japan; [2]Laboratory of Physiology, Atmosphere and Ocean Research Institute, University of Tokyo, Chiba, Japan; [3]Graduate School of Fisheries Science, Hokkaido University, Hakodate, Japan; [4]Faculty of Science, Hokkaido University, Sapporo, Japan; [5]Division of Nephrology, Department of Medicine, University of California, San Diego, CA, USA; [6]Center for Academic Research and Training in Anthropogeny (CARTA), University of California, San Diego, CA, USA

*Correspondence:

ykatsu@sci.hokudai.ac.jp

mbaker@ucsd.edu

**Abstract**.
We investigated progestin and corticosteroid activation of the progesterone receptor (PR) from elephant shark (*Callorhinchus milii*), a cartilaginous fish belonging to the oldest group of jawed vertebrates.   Comparison with human PR experiments provides insights into the evolution of steroid activation of human PR.   At 1 nM steroid, elephant shark PR is activated by progesterone, 17-hydroxy-progesterone, 20β-hydroxy-progesterone, 11-deoxycorticosterone (21-hydroxyprogesterone) and 11-deoxycortisol.   At 1 nM steroid, human PR is activated only by progesterone and 11-deoxycorticosterone indicating increased specificity for progestins and corticosteroids during the evolution of human PR.   RU486, an important clinical antagonist of human PR, did not inhibit progesterone activation of elephant shark PR.   Cys-528 in elephant shark PR corresponds to Gly-722 in human PR, which is essential for RU486 inhibition of human PR.   Confirming the importance of this site on elephant shark PR, RU486 inhibited progesterone activation of the Cys528Gly mutant PR.

There also was a decline in activation of elephant shark Cys528Gly PR by 11-deoxycortisol, 17-hydroxy-progesterone and 20β-hydroxy-progesterone and an increase in activation of human Gly722Cys PR by 11-deoxycortisol and decreased activation by corticosterone.   One or more of these changes may have selected for the mutation corresponding to human glycine-722 PR that first evolved in platypus PR, a basal mammal.

**Running title:** Evolution of steroid activation of elephant shark PR

**Introduction**

The progesterone receptor (PR) receptor belongs to the nuclear receptor family, a diverse group of transcription factors that also includes the glucocorticoid receptor (GR), mineralocorticoid receptor, androgen receptor (AR), and estrogen receptor (ER) (1–3).   In humans, the progesterone receptor (PR) mediates progesterone regulation of female reproductive physiology in the uterus and mammary gland, including fertilization, maintenance of pregnancy and preparation of the endometrium for implantation and parturition (4–7).   Moreover, progesterone has important physiological actions in males, including in the prostate and testes (8–11).   Further, progesterone activates the PR in the brain, bone, thymus, lung and vasculature in females and males (12,13).   Thus, progesterone is a steroid with diverse physiological activities in many organs in females and males.

Although activation by progesterone of the PR in chickens (4,14) , humans (15) , and zebrafish (16,17) has been examined, steroid activation of a PR in the more basal cartilaginous fish lineage has not been fully investigated.   To remedy this omission, we studied the activation by a panel of progestins and corticosteroids (Figure 1) of the PR from the elephant shark (*Callorhinchus milii*), a cartilaginous fish belonging to the oldest group of jawed vertebrates, which diverged about 450 million years ago from bony vertebrates (18,19).

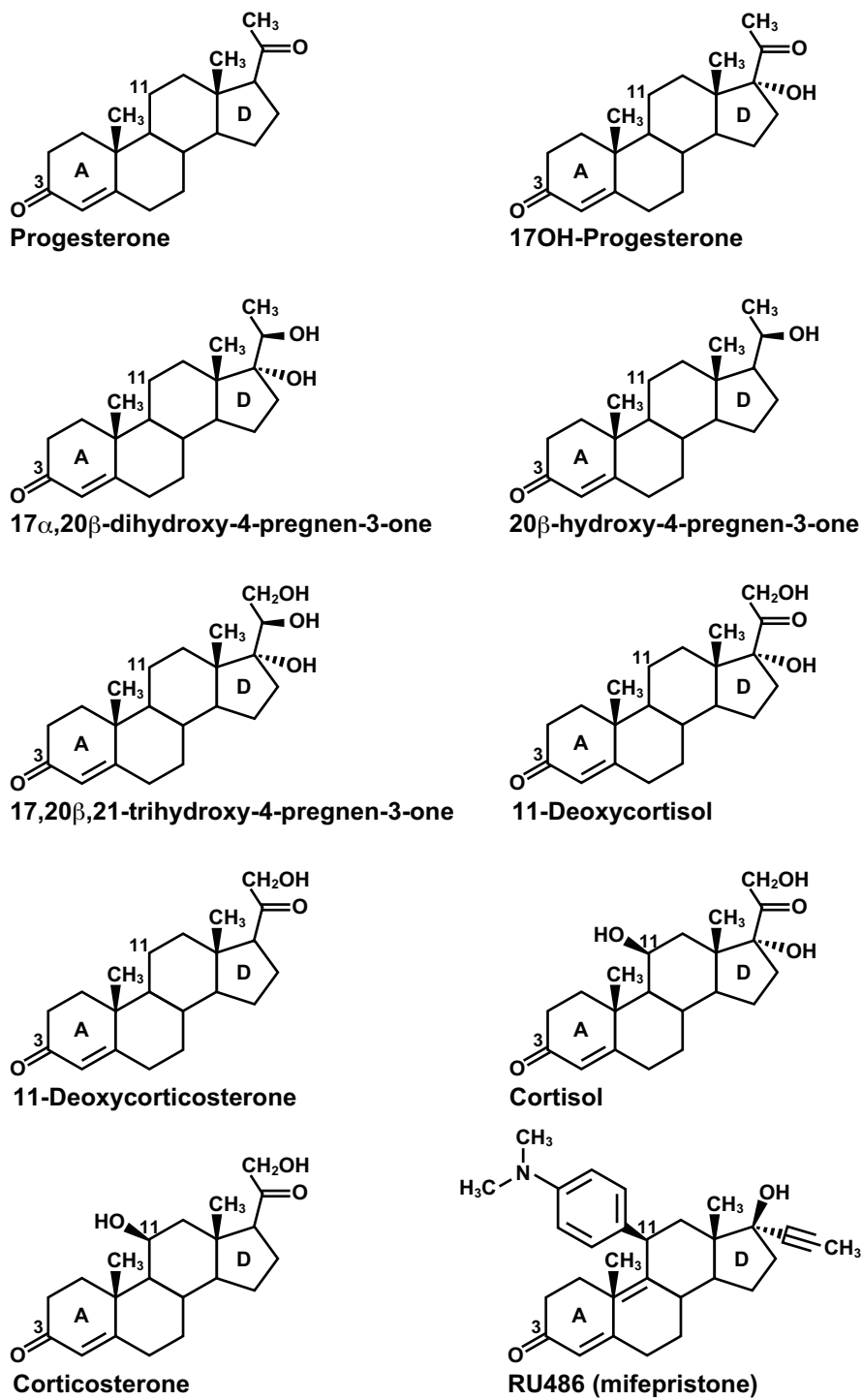

**Figure 1. Structures of corticosteroids and progestins.**

Progesterone is female reproductive steroid that also is important in male physiology (4,11).

17,20β-dihydroxy-progesterone is a maturation inducing hormone of teleost fish (20–22).

17,20β,21-trihydroxy-progesterone is a major ovarian steroid produced by the teleost fish

(23).  Cortisol, corticosterone and 11-deoxycortisol are physiological glucocorticoids in terrestrial vertebrates and ray-finned fish (24,25).   11-deoxycorticosterone is a mineralocorticoid (25–28).   RU486 is an antagonist of human PR (29,30).

Elephant shark PR is an attractive receptor to investigate the ancestral regulation of steroid-mediated PR transcription because, in addition to its phylogenetic position as a member of the oldest lineage of jawed vertebrates, genomic analyses reveal that elephant shark genes are evolving slowly (19), making studies of its PR useful for studying ancestral proteins, including the PR, for comparison for similarities and differences with human PR to elucidate the evolution of steroid specificity for the PR in terrestrial vertebrates (19,31,32). In addition, we were interested in the response of elephant shark PR to RU486 (Mifepristone), which is an antagonist for the human PR (29,30,33) and also a potential anticancer drug for treating progesterone-dependent breast cancer (34).

We find that elephant shark PR is activated by progesterone, 17-hydroxy-progesterone, 20β-hydroxy-progesterone, 17,20β-dihydroxy-progesterone, corticosterone, 11-deoxycorticosterone (21-hydroxy-progesterone) and 11-deoxycortisol.   In contrast human PR is activated only by progesterone, 20β-hydroxy-progesterone, 11-deoxycorticosterone and corticosterone, indicating that human PR has increased specificity for progestins and corticosteroids.

We also find that RU486 does not inhibit progesterone activation of elephant shark PR.   We show that this is due to cysteine-528 in elephant shark PR, which corresponds to glycine-722 on human PR, an amino acid that Benhamou et al. (35) reported was essential for antagonist activity of RU486.   They found that mutation of glycine-722 to cysteine abolished RU486 inhibition of progesterone activation of human PR.   Analyses of vertebrate PRs reveals that an ancestor of human PR-Gly722 first appeared in platypus, a basal mammal (31).

To search for functional changes in human PR that correlate with the evolution of RU486 antagonist activity, we constructed the elephant shark PR-Gly528 mutant and the human PR-Cys722 mutant and studied their activation by several steroids.   Elephant shark PR-Gly528 had a weaker response to 11-deoxycortisol, 17-hydroxy-progesterone and 20β-hydroxy-progesterone.   Human PR-Cys722 displayed increased activation by 11-deoxycortisol and decreased activation by corticosterone.   An altered response to one or more of these steroids may have been selective for the evolution of an ancestor of glycine-722 in a PR in an ancestral platypus at the base of the mammalian line.

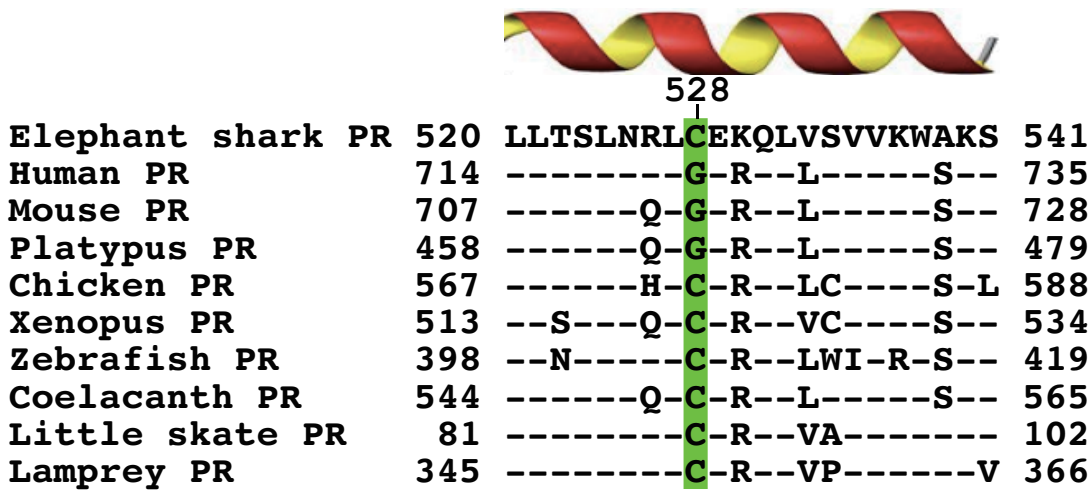

**Figure 2. Alignment of a helix, containing a key amino acid necessary for RU486 inhibition of human PR and activation of elephant shark PR.**
Alignment of α-helix-3 in human PR, containing Gly-722 that is essential for RU486 inhibition of progesterone activation of human PR (35), the PR in elephant shark and other selected vertebrates. RU486 activates elephant shark PR, which contains cysteine-528 corresponding to human PR Gly-722. A glycine first appears in this position in platypus PR, a basal mammal. Amino acids that are identical to amino acids in elephant shark PR are denoted by (-).

Interestingly, 1 nM 17,20β-dihydroxy-progesterone, which is the physiological steroid for the PR in zebrafish (17,27) and other teleosts (20–22,36,37) had less than 10% of the activity of 1 nM progesterone for elephant shark PR suggesting that a role for 17,20β-dihydroxy-progesterone, instead of progesterone, as a ligand for fish PR evolved later in a ray-finned fish (36).

## Materials and Methods
### Chemical reagents

Cortisol, corticosterone, 11-deoxycorticosterone, 11-deoxycortisol, progesterone, 17α-hydroxy-progesterone, 17,20β,21-tri-hydroxy-progesterone, 20β-hydroxy-progesterone, and 17,20β-dihydroxy-progesterone were purchased from Sigma-Aldrich. RU486 was purchased from Cayman Chemical. For reporter gene assays, all hormones were dissolved in dimethyl-sulfoxide (DMSO); the final DMSO concentration in the culture medium did not exceed 0.1%.

**Construction of plasmid vectors**

The full-length PRs were amplified by PCR with KOD DNA polymerase. The PCR products were gel-purified and ligated into pcDNA3.1 vector (Invitrogen). Site-directed mutagenesis was performed using KOD-Plus-mutagenesis kit (TOYOBO). All cloned DNA sequences were verified by sequencing.

**Transactivation assay and statistical methods**

Transfection and reporter assays were carried out in HEK293 cells, as described previously (38,39). All experiments were performed in triplicate. The values shown are mean ± SEM from three separate experiments, and dose-response data, which were used to calculate the half maximal response (EC50) for each steroid, were analyzed using GraphPad Prism. Comparisons between two groups were performed using paired $t$-test. $P < 0.05$ was considered statistically significant. The use of HEK293 cells and an assay temperature of 37C does not replicate the physiological environment of elephant sharks. Nevertheless, studies with HEK293 cells and other mammalian cell lines have proven useful for other studies of transcriptional activation by steroids of steroid hormone receptors from non-mammalian species (39–41).

**Results**
**Transcriptional activation of full-length elephant shark PR by progestins and corticosteroids.**

We screened a panel of steroids for transcriptional activation of full-length elephant shark and human PRs using HEK293 cells. At 10 nM, progesterone, 17-hydroxy-progesterone, 17, 20β-dihydroxy-progesterone, a fish maturation hormone, and 20β-hydroxy-progesterone activated elephant shark PR (Figure 3A). At 10 nM, 11-deoxycorticosterone (21-hydroxyprogesterone), 11-deoxycortisol and corticosterone activated elephant shark PR (Figure 3C) indicating that elephant shark PR responds to corticosteroids.

At 10 nM steroid, human PR responded only to progesterone and 20β-hydroxy-progesterone and not to the other progestins (Figure 3B). 10 nM 11-deoxycorticosterone and corticosterone activated human PR (Figure 3D).

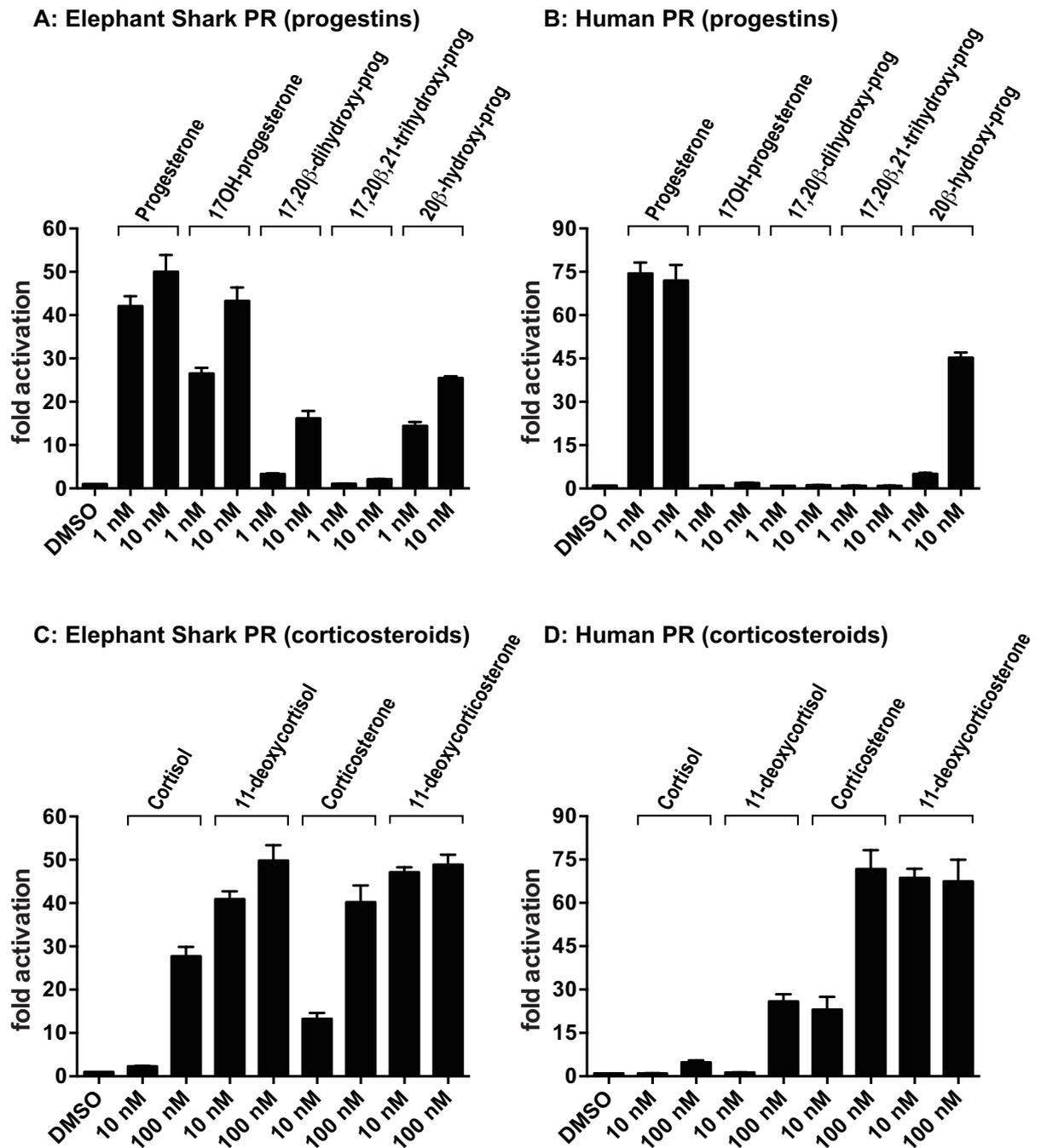

**Figure 3. Transcriptional activation of elephant shark PR by progestins and corticosteroids.**

Elephant shark PR (A and C) and human PR (B and D) were expressed in HEK293 cells with an MMTV-luciferase reporter. Cells were treated with 1 and 10 nM progestins (progesterone, 17OH-progesterone, 17,20β-dihydroxy-progesterone, 17,20β,21-trihydroxy-progesterone and 20β-OH-progesterone), and 10 and 100 nM corticosteroids (cortisol, 11-deoxycortisol, corticosterone, 11-deoxycorticosterone), or vehicle alone (DMSO). Results are expressed as means ± SEM, n=3.   Y-axis indicates fold-activation compared to the activity by vehicle (DMSO) alone as 1.

**Concentration-dependent activation by progestins and corticosteroids of elephant shark PR and human PR**.

To gain a quantitative measure of progestin and corticosteroid activation of elephant shark PR and human PR, we determined the concentration dependence of transcriptional activation by progestins and corticosteroids of elephant shark PR (Figure 4A, C) and for comparison activation of human PR (Figure 4B, D).   Progesterone, 17OH-progesterone, 17,20β-dihydroxy-progesterone, 20β-OH-progesterone, 11-deoxycortisol, corticosterone and 11-deoxycorticosterone activated elephant shark PR, while human PR was stimulated only by progesterone, 20β-OH-progesterone, corticosterone and 11-deoxycorticosterone, a more limited number of steroids.

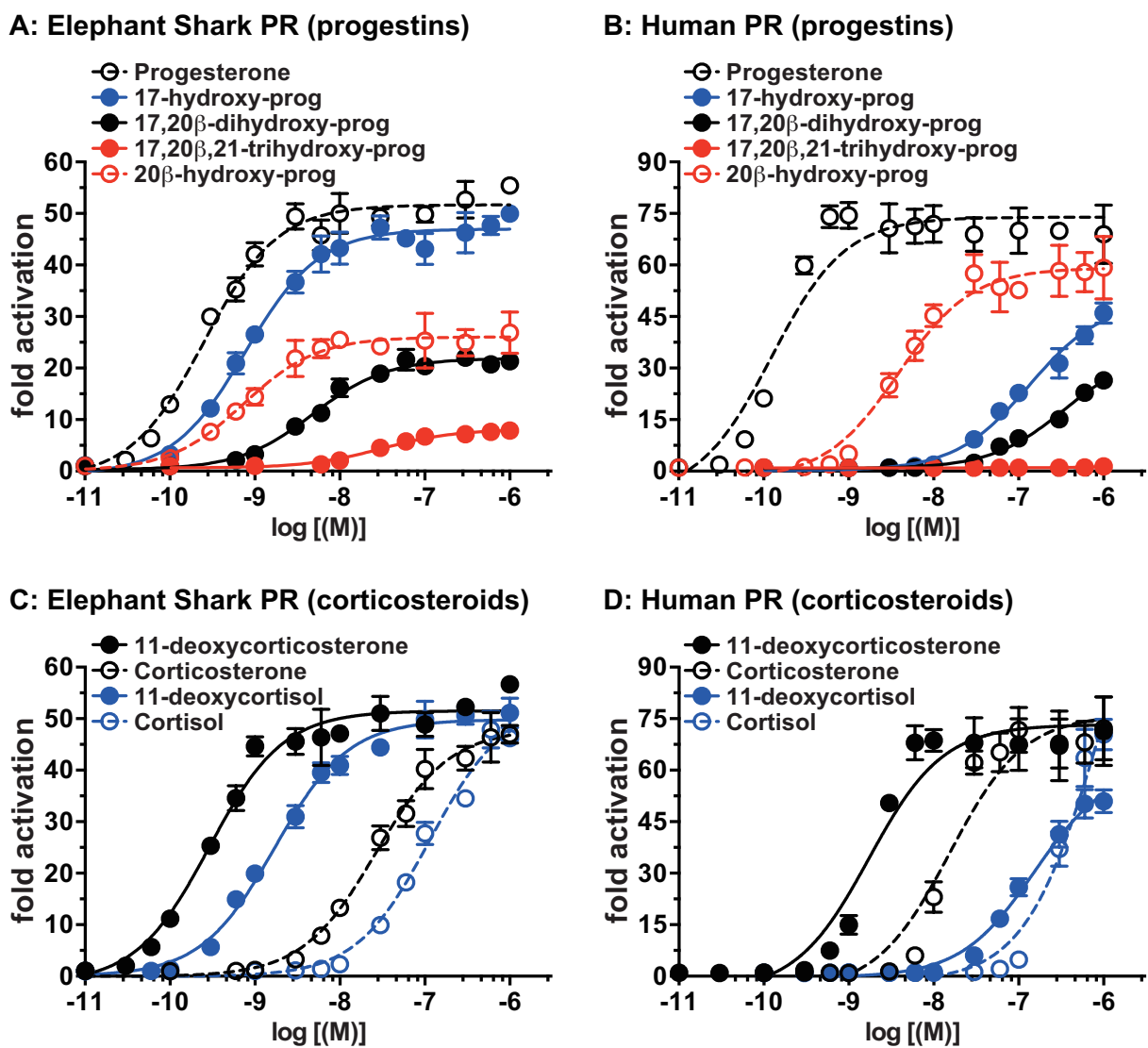

**Figure 4. Concentration-dependent transcriptional activation of elephant shark and human PRs by corticosteroids and progestins.**
Elephant shark PR (A and C) and human PR (B and D) were expressed in HEK293 cells with an MMTV-luciferase reporter.   Cells were treated with increasing concentrations of corticosteroids (A and B), progestins (C and D) or vehicle alone (DMSO).   Y-axis indicates fold-activation compared to the activity by vehicle (DMSO) alone as 1.

Table 1 summarizes the EC50s of progestins and corticosteroids for elephant shark PR and human PR.   We find that elephant shark PR has low EC50s for progesterone (0.25 nM), 17-OH-progesterone (0.77 nM), 20β-OH-progesterone (0.73 nM), 17α,20β-OH-progesterone, 11-deoxycortisol (1.7 nM) and 11-deoxycorticosterone (0.29 nM). In contrast, human PR has low EC50s for progesterone (0.12 nM), 20β-OH-progesterone (3.8 nM) and 11-deoxycorticosterone (1.7 nM).

Table 1. EC50 values for steroid activation of elephant shark PR and human PR

|  | Prog | 17OH-Prog | 17α,20β-OH-Prog | 17α,20β,21-tri-OH-Prog | 20β-OH-Prog |
|---|---|---|---|---|---|
|  | EC50 (nM) | EC50 (nM) | EC50 (nM) | EC50 (nM) | EC50 (nM) |
| Elephant shark PR | 0.25 | 0.77 | 4.8 | 30.9 | 0.73 |
| 95% confidence intervals | 0.2-0.33 | 0.58-1.0 | 3.6-6.6 | 23.3-41.1 | 0.52-1.0 |
| Human PR | 0.12 | 132 | 344 | - | 3.8 |
| 95% confidence intervals | 0.079-0.19 | 94.7-183 | 237-498 | - | 2.9-5.1 |

|  | Cortisol | 11-deoxycortisol | Corticosterone | DOC |
|---|---|---|---|---|
|  | EC50 (nM) | EC50 (nM) | EC50 (nM) | EC50 (nM) |
| Elephant shark PR | 114 | 1.7 | 26.5 | 0.29 |
| 95% confidence intervals | 88.1-148 | 1.4-2.1 | 20.0-35.2 | 0.22-0.38 |
| Human PR | 882 | 154 | 15.4 | 1.7 |
| 95% confidence intervals | 424-1835 | 115-205 | 9.5-24.9 | 1.1-2.8 |

Prog = progesterone, 17OH-Prog = 17OH-progesterone, 17α,20β-OH-Prog= 17α,20β-dihydroxy-progesterone, 17α,20β,21-tri-OH-Prog = 17α,20β,21-trihydroxy-progesterone, 20β-OH-Prog = 20β-hydroxy-progesterone, DOC =11-deoxycorticosterone,

**RU486 does not inhibit transactivation of elephant shark PR**

Activation of human PR by progesterone is inhibited by RU486 (29,30,33). Indeed, at 0.1 nM and 1 nM, RU486 inhibits of activation by 1 nM progesterone of human PR (Figure 5A). Benhamou et al. (35) reported that Gly-722 in human PR, is essential for the inhibition of progesterone activation of human PR by of RU486. We confirm that RU486 does not inhibit the human PR Cys722 mutant (Figure 5C).

Cys-528 of elephant shark PR corresponds to Gly-722 of human PR, which predicts that RU486 would not inhibit progesterone activation of wild type elephant shark PR, and indeed, as shown in Figure 5B, 1 nM progesterone activation of elephant shark PR was not inhibited by 100 nM RU486.   As expected, RU486 inhibits progesterone activation of elephant shark PR Gly528 (Figure 5D).

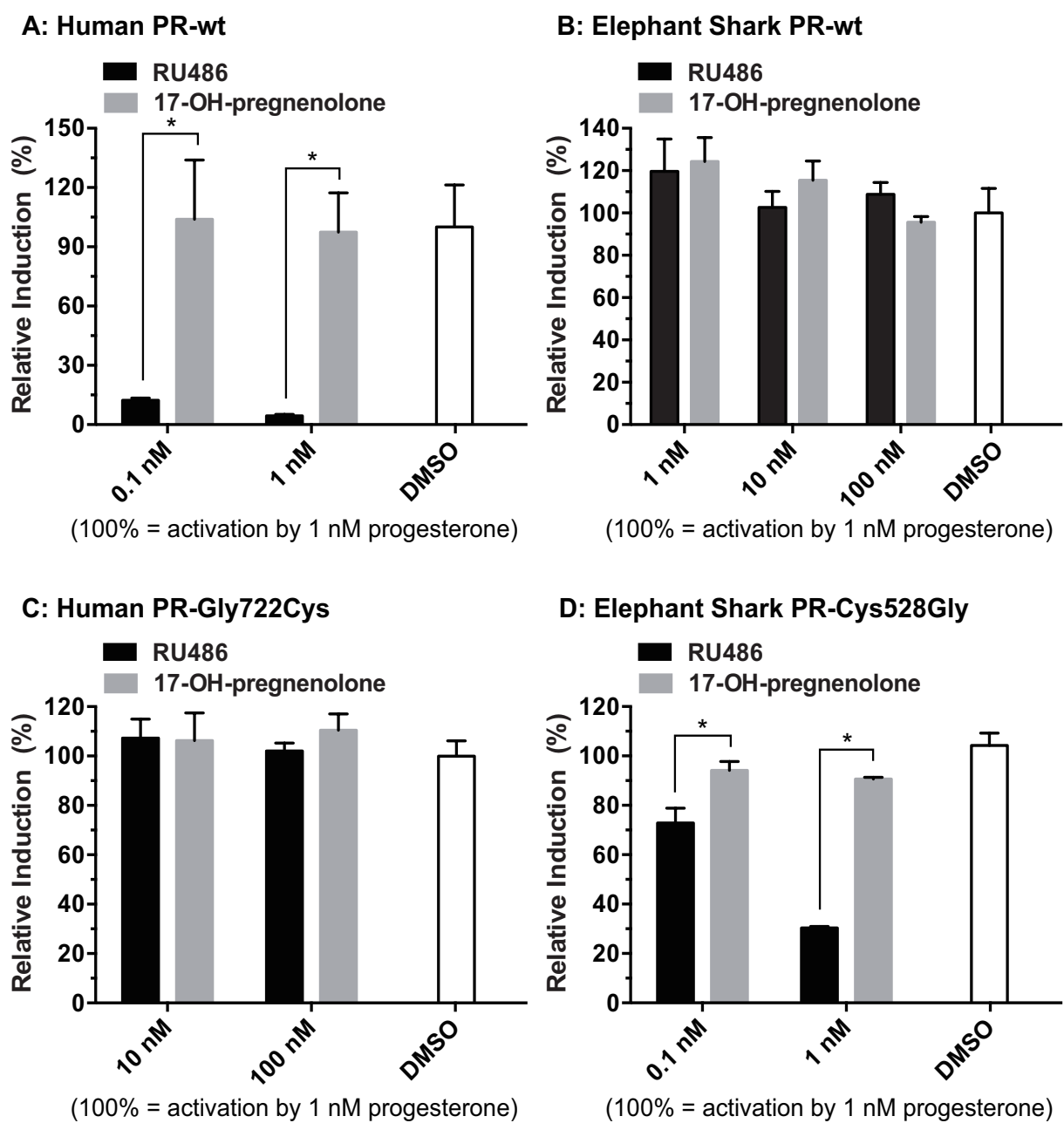

**Figure 5. Effect of RU486 for Prog-induced activation of PR.**
Wild-type of human PR (A) or elephant shark PR (B) was expressed in HEK293 cells with an MMTV-luciferase reporter.   Cells with human PR were treated with 1 nM progesterone and either 0.1 nM or 1 nM RU486, 17OH- pregnenolone or DMSO.   Cells with elephant shark

PR were treated with 1 nM progesterone and with either 1 nM, 10 nM or 100 nM RU486, 17OH-pregnenolone or DMSO.   Human PR-Gly722Cys (C) or elephant shark PR-Cys528Gly (D) was expressed in HEK293 cells with an MMTV-luciferase reporter.   Cells with human PR-Gly722Cys were treated with 1 nM progesterone and with either 10 nM or 100 nM RU486, 17OH-pregnenolone or DMSO.   Cells with elephant shark PR-Cys528Gly were treated with 1 nM progesterone and with either 0.1 nM or 1 nM RU486 or 17OH-Pregnenolone   Relative inductions were normalized between 0 and 100%, where 0 and 100 were defined as the bottom and tip value in vehicle-treated and 1 nM progesterone treated, respectively.   Results are expressed as means ± SEM, n=3.   * $P < 0.05$ compared with vehicle treatment (student's *t*-test).

**Steroid activation of human PR Gly722Cys and elephant shark PR Cys528Gly.**

To search for a biological basis for the functional changes in human PR due to Gly-722 in human PR we constructed a human PR-Cys722 mutant and an elephant shark PR-Gly528 mutant studied their activation by various progestins and corticosteroids (Figure 6). Elephant shark PR-Gly528 had a weaker response to 11-deoxycortisol, 17-hydroxy-progesterone and 20β-hydroxy-progesterone.   Human PR-Cys722 displayed increased activation by 11-deoxycortisol and decreased activation by corticosterone.   One or more of these changes may have been selective for the evolution an ancestor of glycine-722 in a PR in an ancestral platypus at the base of the mammalian line.

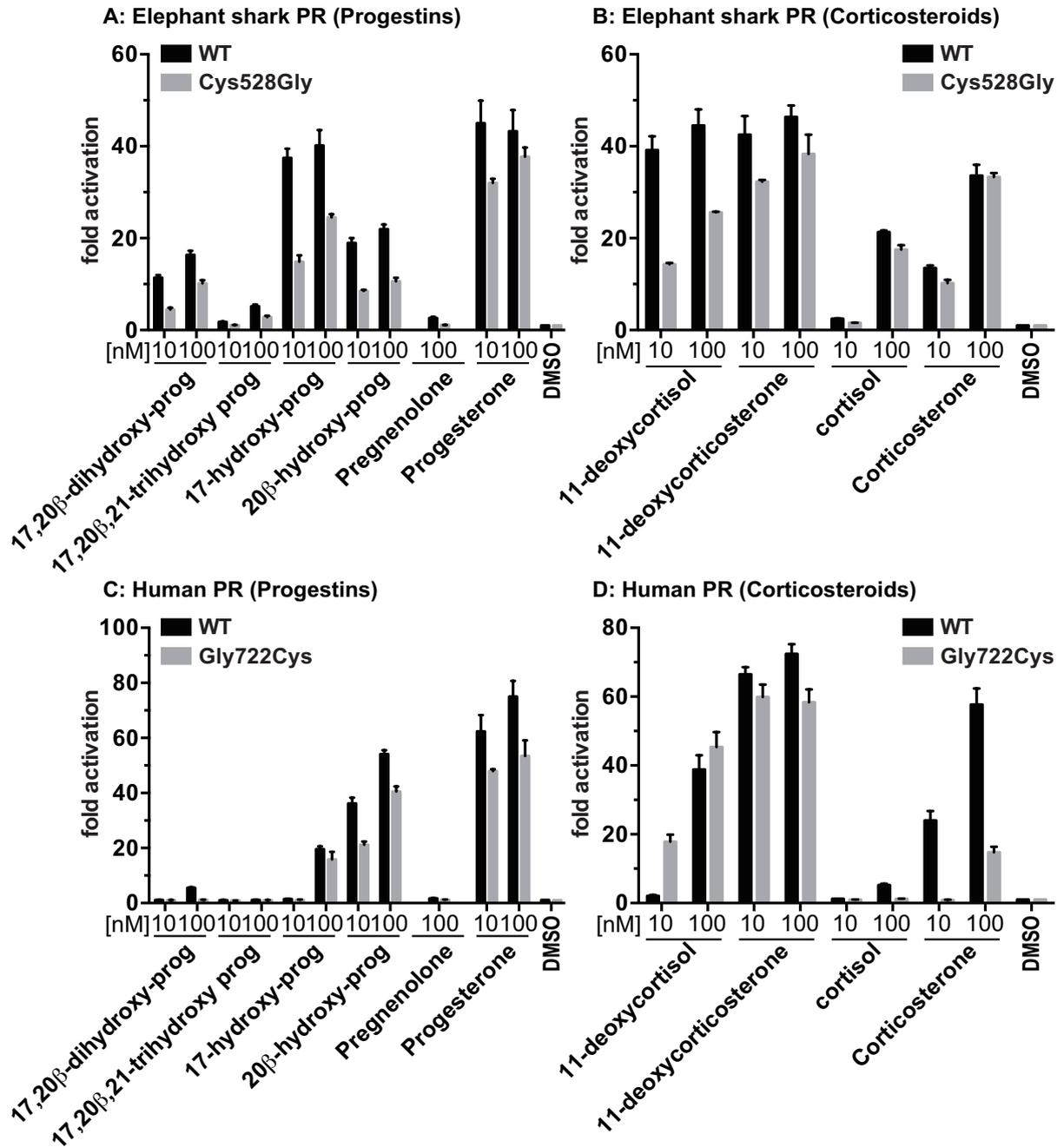

**Figure 6. Ligand-dependent transcriptional activation of elephant shark PR-C528G and human PR-G722C.**

Elephant shark PR (A and C), and human PR (B and D) were expressed in HEK293 cells with an MMTV-luciferase reporter. Cells transfected with PRs were treated with increasing concentrations of Prog or vehicle alone (DMSO) (A and B). Cells were treated with 10 nM progestins (Prog, 17OH-Progesterone, pregnenolone 17,20β-dihydroxy-progesterone, 17,20β,21-trihydroxy-progesterone, 20β-OH-progesterone), corticosteroids (cortisol, 11-deoxycortisol, corticosterone, 11-deoxycorticosterone), or vehicle alone (DMSO) (C and D).

Results are expressed as means ± SEM, n=4. Y-axis indicates fold-activation compared to the activity of control vector with vehicle (DMSO) alone as 1.

**Discussion**

An ortholog of human PR along with the corticoid receptor (CR), the ancestor of the MR and GR, first appears in the more ancient cyclostomes (jawless fish), which has descendants in modern lamprey and hagfish (3,27,42). It is in cartilaginous fish that distinct orthologs of human MR and GR first appear (43–45) along with the beginning of the evolution of differences in the responses of the PR, MR and GR for progestins and corticosteroids that appear in terrestrial vertebrates and ray-finned fish (16,17,24,40,41,43–51).

Here we report that elephant shark PR has a strong response to progesterone (EC50 0.25 nM) and 17-OH-progesterone (EC50 0.77 nM), as well as to 11-deoxycorticosterone (EC50 0.29 nM) (Figure 4, Table 1), a corticosteroid with close structural similarity to progesterone (Figure 1). Elephant shark PR also is activated by 20β-OH-progesterone (EC50 0.73 nM) and 17,20β-dihydroxy-progesterone (EC50 4.8 nM) and 11-deoxycortisol (EC50 1.7 nM). This contrasts with selectivity of human PR, which has a strong response to progesterone (EC50 0.12 nM) and 11-deoxycorticosterone, (EC50 1.7 nM) and weak response to corticosterone, (EC50 15.4 nM). The advantage this selectivity of human PR is not known.

The evolution of the response of human PR to RU486 provides intriguing clues because RU486 is not a physiological ligand. The glycine-722 in human PR (Figure 2) that confers antagonist activity for RU486 towards human PR first appears in platypus, a basal mammal (Figure 2). Human PR with a Cys722 mutation, to mimic the ancestral PR, had increased activation by 11-deoxycortisol and decreased activation by corticosterone.

The lower activation of elephant shark PR by 17,20β-dihydroxy-progesterone compared to progesterone is intriguing because in zebrafish (16,17), this is reversed with higher activity for 17,20β-dihydroxy-progesterone compared to progesterone indicating that during the evolution of ray-finned fish the response of the PR to progesterone diminished and the response to 17,20β-dihydroxy-progesterone increased(21,22,36,37).

**Funding, Contributions and Competing Interests.**
**Funding:** This work was supported by Grants-in-Aid for Scientific Research from the Ministry of Education, Culture, Sports, Science and Technology of Japan (19K067309 to Y.K.), and the Takeda Science Foundation (to Y.K.).   M.E.B. was supported by Research fund #3096.

**Author contributions:** X.L. and Y.K. carried out the research and analyzed data.   W.T. and S.H. aided in the collection of animals.   S.I. provided steroids used in this study.   Y.K. and M.E.B. conceived and designed the experiments.   X.L., Y.K. and M.E.B. wrote the paper. All authors gave final approval for publication.

**Competing Interests:** We have no competing interests.

**References**

1. Evans RM. The steroid and thyroid hormone receptor superfamily. Science. 1988;240(4854):889-895. doi:10.1126/science.3283939.

2. Baker ME. Steroid receptors and vertebrate evolution. Mol Cell Endocrinol. 2019;496:110526. doi:10.1016/j.mce.2019.110526.

3. Bridgham JT, Eick GN, Larroux C, et al. Protein evolution by molecular tinkering: diversification of the nuclear receptor superfamily from a ligand-dependent ancestor. PLoS Biol. 2010;8(10):e1000497. Published 2010 Oct 5. doi:10.1371/journal.pbio.1000497.

4. Conneely OM, Mulac-Jericevic B, DeMayo F, Lydon JP, O'Malley BW. Reproductive functions of progesterone receptors. Recent Prog Horm Res. 2002;57:339-355. doi:10.1210/rp.57.1.339.

5. Graham JD, Clarke CL. Physiological action of progesterone in target tissues. Endocr Rev. 1997;18(4):502-519. doi:10.1210/edrv.18.4.0308.

6. Obr AE, Edwards DP. The biology of progesterone receptor in the normal mammary gland and in breast cancer. Mol Cell Endocrinol. 2012;357(1-2):4-17. doi:10.1016/j.mce.2011.10.030.

7. Smith R. Parturition. N Engl J Med. 2007;356(3):271-283. doi:10.1056/NEJMra061360.

8. Aquila S, De Amicis F. Steroid receptors and their ligands: effects on male gamete functions. Exp Cell Res. 2014;328(2):303-313. doi:10.1016/j.yexcr.2014.07.015.

9. Luetjens CM, Didolkar A, Kliesch S, et al. Tissue expression of the nuclear progesterone receptor in male non-human primates and men. J Endocrinol. 2006;189(3):529-539. doi:10.1677/joe.1.06348.

10. Grindstad T, Richardsen E, Andersen S, et al. Progesterone Receptors in Prostate Cancer: Progesterone receptor B is the isoform associated with disease progression. Sci Rep. 2018;8(1):11358. Published 2018 Jul 27. doi:10.1038/s41598-018-29520-5.

11. Publicover S, Barratt C. Reproductive biology: Progesterone's gateway into sperm [published correction appears in Nature. 2011 Mar 31;471(7340):589]. Nature. 2011;471(7338):313-314. doi:10.1038/471313a.

12. Crews D. Evolution of neuroendocrine mechanisms that regulate sexual behavior. Trends Endocrinol Metab. 2005;16(8):354-361. doi:10.1016/j.tem.2005.08.007.

13. Wagner CK. The many faces of progesterone: a role in adult and developing male brain. Front Neuroendocrinol. 2006;27(3):340-359. doi:10.1016/j.yfrne.2006.07.003.

14. Gronemeyer H, Turcotte B, Quirin-Stricker C, Bocquel MT, Meyer ME, Krozowski Z, Jeltsch JM, Lerouge T, Garnier JM, Chambon P. The chicken progesterone receptor: sequence, expression and functional analysis. EMBO J. 1987 Dec 20;6(13):3985-94. PMID: 3443098; PMCID: PMC553878.

15. Grimm SL, Hartig SM, Edwards DP. Progesterone Receptor Signaling Mechanisms. J Mol Biol. 2016 Sep 25;428(19):3831-49. doi: 10.1016/j.jmb.2016.06.020. Epub 2016 Jul 2. PMID: 27380738.

16. Chen SX, Bogerd J, García-López A, et al. Molecular cloning and functional characterization of a zebrafish nuclear progesterone receptor. Biol Reprod. 2010;82(1):171-181. doi:10.1095/biolreprod.109.077644.

17. Hanna RN, Daly SC, Pang Y, et al. Characterization and expression of the nuclear progestin receptor in zebrafish gonads and brain. Biol Reprod. 2010;82(1):112-122. doi:10.1095/biolreprod.109.078527.

18. Hara Y, Yamaguchi K, Onimaru K, Kadota M, Koyanagi M, Keeley SD, Tatsumi K, Tanaka K, Motone F, Kageyama Y, Nozu R, Adachi N, Nishimura O, Nakagawa R, Tanegashima C, Kiyatake I, Matsumoto R, Murakumo K, Nishida K, Terakita A,

Kuratani S, Sato K, Hyodo S, Kuraku S. Shark genomes provide insights into elasmobranch evolution and the origin of vertebrates. Nat Ecol Evol. 2018 Nov;2(11):1761-1771. doi: 10.1038/s41559-018-0673-5. Epub 2018 Oct 8. PMID: 30297745.

19. Venkatesh B, Lee AP, Ravi V, et al. Elephant shark genome provides unique insights into gnathostome evolution [published correction appears in Nature. 2014 Sep 25;513(7519):574]. Nature. 2014;505(7482):174-179. doi:10.1038/nature12826.

20. Nagahama Y, Adachi S. Identification of maturation-inducing steroid in a teleost, the amago salmon (Oncorhynchus rhodurus). Dev Biol. 1985 Jun;109(2):428-35. doi: 10.1016/0012-1606(85)90469-5. PMID: 3996758.

21. Ikeuchi T, Todo T, Kobayashi T, Nagahama Y. A novel progestogen receptor subtype in the Japanese eel, Anguilla japonica. FEBS Lett. 2002;510(1-2):77-82. doi:10.1016/s0014-5793(01)03220-3.

22. Scott AP, Sumpter JP, Stacey N. The role of the maturation-inducing steroid, 17,20beta-dihydroxypregn-4-en-3-one, in male fishes: a review. J Fish Biol. 2010;76(1):183-224. doi:10.1111/j.1095-8649.2009.02483.x.

23. Trant JM, Thomas P, Shackleton CH. Identification of 17 alpha,20 beta,21-trihydroxy-4-pregnen-3-one as the major ovarian steroid produced by the teleost Micropogonias undulatus during final oocyte maturation. Steroids. 1986 Feb-Mar;47(2-3):89-99. doi: 10.1016/0039-128x(86)90081-4. PMID: 3564088.

24. Sturm A, Bury N, Dengreville L, et al. 11-deoxycorticosterone is a potent agonist of the rainbow trout (Oncorhynchus mykiss) mineralocorticoid receptor. Endocrinology. 2005;146(1):47-55. doi:10.1210/en.2004-0128.

25. Baker ME, Funder JW, Kattoula SR. Evolution of hormone selectivity in glucocorticoid and mineralocorticoid receptors [published correction appears in J Steroid Biochem Mol Biol. 2014 Jan;139:104]. J Steroid Biochem Mol Biol. 2013;137:57-70. doi:10.1016/j.jsbmb.2013.07.009.

26. Hawkins UA, Gomez-Sanchez EP, Gomez-Sanchez CM, Gomez-Sanchez CE. The ubiquitous mineralocorticoid receptor: clinical implications. Curr Hypertens Rep. 2012;14(6):573-580. doi:10.1007/s11906-012-0297-0.

27. Rossier BC, Baker ME, Studer RA. Epithelial sodium transport and its control by aldosterone: the story of our internal environment revisited. Physiol Rev. 2015;95(1):297-340. doi:10.1152/physrev.00011.2014.

28. Fuller PJ, Yao Y, Yang J, Young MJ. Mechanisms of ligand specificity of the mineralocorticoid receptor. J Endocrinol. 2012 Apr;213(1):15-24. doi: 10.1530/JOE-11-0372. Epub 2011 Dec 12. PMID: 22159507.

29. Baulieu EE. Contragestion and other clinical applications of RU 486, an antiprogesterone at the receptor. Science. 1989;245(4924):1351-1357. doi:10.1126/science.2781282.

30. Cadepond F, Ulmann A, Baulieu EE. RU486 (mifepristone): mechanisms of action and clinical uses. Annu Rev Med. 1997;48:129-156. doi:10.1146/annurev.med.48.1.129.

31. Baker ME, Uh KY. Evolutionary analysis of the segment from helix 3 through helix 5 in vertebrate progesterone receptors. J Steroid Biochem Mol Biol. 2012 Oct;132(1-2):32-40. doi: 10.1016/j.jsbmb.2012.04.007. Epub 2012 Apr 30. PMID: 22575083.

32. Inoue JG, Miya M, Lam K, et al. Evolutionary origin and phylogeny of the modern holocephalans (Chondrichthyes: Chimaeriformes): a mitogenomic perspective. Mol Biol Evol. 2010;27(11):2576-2586. doi:10.1093/molbev/msq147.

33. Mao J, Regelson W, Kalimi M. Molecular mechanism of RU 486 action: a review. Mol Cell Biochem. 1992 Jan 15;109(1):1-8. doi: 10.1007/BF00230867. PMID: 1614417.

34. Klijn JG, Setyono-Han B, Foekens JA. Progesterone antagonists and progesterone receptor modulators in the treatment of breast cancer. Steroids. 2000 Oct-Nov;65(10-11):825-30. doi: 10.1016/s0039-128x(00)00195-1. PMID: 11108894.

35. Benhamou B, Garcia T, Lerouge T, Vergezac A, Gofflo D, Bigogne C, Chambon P, Gronemeyer H. A single amino acid that determines the sensitivity of progesterone receptors to RU486. Science. 1992 Jan 10;255(5041):206-9. doi: 10.1126/science.1372753. PMID: 1372753.

36. Todo T, Ikeuchi T, Kobayashi T, Kajiura-Kobayashi H, Suzuki K, Yoshikuni M, Yamauchi K, Nagahama Y. Characterization of a testicular 17alpha, 20beta-dihydroxy-4-pregnen-3-one (a spermiation-inducing steroid in fish) receptor from a teleost,

Japanese eel (Anguilla japonica). FEBS Lett. 2000 Jan 7;465(1):12-7. doi: 10.1016/s0014-5793(99)01714-7. PMID: 10620698.

37. Canario AV, Scott AP. Structure-activity relationships of C21 steroids in an in vitro oocyte maturation bioassay in rainbow trout, Salmo gairdneri. Gen Comp Endocrinol. 1988 Aug;71(2):338-48. doi: 10.1016/0016-6480(88)90262-6. PMID: 3203879.

38. Oka K, Hoang A, Okada D, Iguchi T, Baker ME, Katsu Y. Allosteric role of the amino-terminal A/B domain on corticosteroid transactivation of gar and human glucocorticoid receptors. J Steroid Biochem Mol Biol. 2015;154:112-119. doi:10.1016/j.jsbmb.2015.07.025.

39. Katsu Y, Oka K, Baker ME. Evolution of human, chicken, alligator, frog, and zebrafish mineralocorticoid receptors: Allosteric influence on steroid specificity. Sci Signal. 2018;11(537):eaao1520. Published 2018 Jul 3. doi:10.1126/scisignal.aao1.

40. Carroll SM, Bridgham JT, Thornton JW. Evolution of hormone signaling in elasmobranchs by exploitation of promiscuous receptors. Mol Biol Evol. 2008;25(12):2643-2652. doi:10.1093/molbev/msn204.

41. Sugimoto A, Oka K, Sato R, Adachi S, Baker ME, Katsu Y. Corticosteroid and progesterone transactivation of mineralocorticoid receptors from Amur sturgeon and tropical gar. Biochem J. 2016;473(20):3655-3665. doi:10.1042/BCJ20160579.

42. Baker ME, Nelson DR, Studer RA. Origin of the response to adrenal and sex steroids: Roles of promiscuity and co-evolution of enzymes and steroid receptors. J Steroid Biochem Mol Biol. 2015;151:12-24. doi:10.1016/j.jsbmb.2014.10.020.

43. Bridgham JT, Carroll SM, Thornton JW. Evolution of hormone-receptor complexity by molecular exploitation. Science. 2006;312(5770):97-101. doi:10.1126/science.1123348.

44. Baker ME, Katsu Y. 30 YEARS OF THE MINERALOCORTICOID RECEPTOR: Evolution of the mineralocorticoid receptor: sequence, structure and function. J Endocrinol. 2017;234(1):T1-T16. doi:10.1530/JOE-16-0661.

45. Baker ME, Katsu Y. Progesterone: An enigmatic ligand for the mineralocorticoid receptor. Biochem Pharmacol. 2020;177:113976. doi:10.1016/j.bcp.2020.113976.

46. Prunet P, Sturm A, Milla S. Multiple corticosteroid receptors in fish: from old ideas to new concepts. Gen Comp Endocrinol. 2006 May 15;147(1):17-23. doi: 10.1016/j.ygcen.2006.01.015. Epub 2006 Mar 20. PMID: 16545810.

47. Sakamoto T, Mori C, Minami S, et al. Corticosteroids stimulate the amphibious behavior in mudskipper: potential role of mineralocorticoid receptors in teleost fish. Physiol Behav. 2011;104(5):923-928. doi:10.1016/j.physbeh.2011.06.002.

48. Takahashi H, Sakamoto T. The role of "mineralocorticoids" in teleost fish: relative importance of glucocorticoid signaling in the osmoregulation and "central" actions of mineralocorticoid receptor. Gen Comp Endocrinol. 2013;181:223-228. doi:10.1016/j.ygcen.2012.11.016.

49. McCormick SD, Bradshaw D. Hormonal control of salt and water balance in vertebrates. Gen Comp Endocrinol. 2006 May 15;147(1):3-8. doi: 10.1016/j.ygcen.2005.12.009. Epub 2006 Feb 2. PMID: 16457828.

50. Katsu Y, Kohno S, Oka K, et al. Transcriptional activation of elephant shark mineralocorticoid receptor by corticosteroids, progesterone, and spironolactone. Sci Signal. 2019;12(584):eaar2668. Published 2019 Jun 4. doi:10.1126/scisignal.aar2668, (n.d.).

51. Fuller PJ, Yao YZ, Jin R, et al. Molecular evolution of the switch for progesterone and spironolactone from mineralocorticoid receptor agonist to antagonist. Proc Natl Acad Sci U S A. 2019;116(37):18578-18583. doi:10.1073/pnas.1903172116, (n.d.).